\documentclass[aps,prb,twocolumn,floatfix,floats,showpacs,superscriptaddress]{revtex4}
\usepackage{graphicx,latexsym}
\usepackage{dcolumn}
\usepackage{amsmath}

\begin{document}
\title{Coherent quantum transport in the presence of  \\
a finite-range transversely polarized time-dependent field}
\author{C. S. Tang}
\affiliation{Physics Division, National Center for Theoretical
        Sciences, P.O.\ Box 2-131, Hsinchu 30013, Taiwan}
\author{C. S. Chu}
\affiliation{Department of Electrophysics, National Chiao Tung
University, Hsinchu 30010, Taiwan}
\begin{abstract}
This work
investigates the quantum transport in a narrow constriction acted
upon by a finite-range transversely polarized time-dependent
electric field.  A generalized scattering-matrix method is developed
that has incorporated a time-dependent mode-matching scheme.  The
transverse field induces coherent inelastic scatterings that include
both intersubband and intersideband transitions.  These scatterings
give rise to the dc conductance $G$ a general suppressed feature
that escalates with the chemical potential.  In addition, particular
suppressed features -- the dip structures -- are found in $G$.
These features are recognized as the quasi-bound-state (QBS)
features that arise from electrons making intersubband transitions
to the vicinity of a subband bottom.  For the case of larger field
intensities, the QBS features that involve more photons are more
evident.  These QBS features are closely associated with the
singular density of states at the subband bottoms. An experimental
setup is proposed for the observation of these features.
\end{abstract}
\pacs{73.23.-b, 72.10.-d, 72.40.+w}
\maketitle

\section{Introduction}

Advances in the epitaxial growth technologies have lead to the
fabrication of high-quality two-dimensional electron gas (2DEG)
systems that are almost defect-free and upon which electronic
nanostructures can be built.  The electron transport properties of
these nanostructures have been studied extensively both experimentally
and theoretically.~\cite{fer97,imr97}  The most studied nanostructure
is the quantum point contact (QPC), due to its simple configuration,
and due also to the significant quantization effects in such systems,
as is shown in the conductance $G$.~\cite{wee88,wha88,wee88b}

\begin{figure}[b]
\includegraphics[width=0.41\textwidth,angle=0]{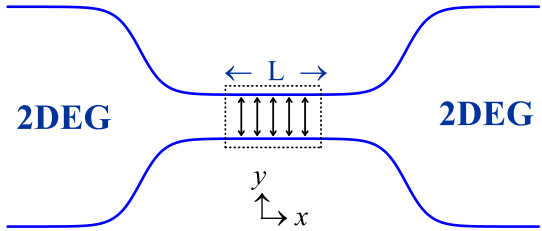}
\caption{ Sketch of the gated QPC which is connected at each end to
a two-dimensional electron gas electrode. The narrow constriction is
acted upon by external transversely polarized time-dependent
electric field within millimeter wave regime. } \label{fig:1}
\end{figure}
These QPC's, when created electrostatically by negatively biasing
a split-gate located on top of a 2DEG,~\cite{wee88,wha88,wee88b}
can be pictured as a narrow constriction  connecting
adiabatically at each end to a 2DEG,~\cite{gla88,gla90} as depicted
in Fig.~\ref{fig:1}. The energy levels in the narrow constriction are
quantized into one-dimensional subbands which density of states (DOS)
is singular near a subband bottom.  This singular DOS was found, in
the presence of an attractive scatterer, to give rise to dip
structures in $G$,~\cite{chu89,bag90,tek91,nix91,lev92,tak92,kun92}
which is associated with the formation of impurity-induced
QBS's~\cite{bag90} formed just beneath a subband bottom.

More recently, attentions have been shifted to QPC's acted upon by
high frequency fields.  These time-dependent fields include
transversely,\,~\cite{hek91,hu93,wys93,wys95,hu96,ala98,%
fed93,jan94,gor94,gri95,maa96,chu96,tag96,tag97} or
longitudinally\,~\cite{fen93,tan99} polarized fields, or simply gate-induced
time-modulated potentials.~\cite{bag92,tan96,tan97}  These studies
focus on coherent inelastic scatterings by assuming the range of
the time-modulated fields to be shorter than the incoherent mean free
path.  A number of interesting transport characteristics were
explored.  In the case of zero source-drain bias, it was demonstrated
that electron pumping could occur when an asymmetric spit-gate is
acted upon by a time-modulated field.\,\cite{hek91}  In the case of a
finite source-drain bias, and the QPC's have varying widths, the
photon-assisted quantum transport characteristics have been
studied.~\cite{hu93,wys93,wys95,hu96,ala98,fed93,jan94,gor94,gri95,maa96}
One might also be prompted by the impurity-induced QBS
features,~\cite{chu89,bag90,tek91,nix91,lev92,tak92,kun92} and ask
whether the QBS's could find their way of manifestation in the time-dependent
transport characteristics in QPC's.  Indeed, earlier studies have
already found QBS features for cases when the QPC's have either a
delta-profile oscillating barrier~\cite{bag92} or transverse field.
\cite{chu96}  In a recent study, we consider the more realistic case
of a finite-range longitudinally polarized field,
\cite{tan99} and we find QBS features that are associated with
electrons making intrasubband and intersideband transitions to the
vicinity of a subband bottom.  Since the transitions allowed depend on
the polarization of the time-modulated field, we opt to investigate
in this paper the QBS features for a finite-range transversely
polarized field.

In theoretical studies of coherent quantum transport in mesoscopic
nanostructures, the transfer-matrix and the scattering-matrix method
are powerful tools. Both the methods enable us to
numerically calculate current transmission
probabilities for arbitrary shaped configurations.
The conductance of the nanostructure can be obtained by
Landauer-B\"uttiker formalism.~\cite{lan57,but86}
However, since the transfer-matrix method fails in systems for
which energy conservation is violated, it is unapt for a
time-dependent external field acting upon the system. Thus,
one has to develop a generalized scattering-matrix method to
calculate numerically the current transmission probability of a
time-modulated mesoscopic nanostructure.

In this paper, we develop a generalized scattering method for
a finite-range transversely polarized time-dependent
electric field acting upon a narrow constriction,
as depicted in Fig.~\ref{fig:1}.
This method enables us to calculate not only the current transmission
and reflection probabilities but also the contributions from each
subband and sideband states yielded by the transversely polarized
electric field. This transverse field
${\bf E}({\bf x},t)={\cal E}(x) \cos(\omega t) \hat{y}$
has a finite longitudinal profile, namely that ${\cal E}(x)$ covers
a length $L$ that excludes the reservoirs.
The finiteness of the electric field in the $x$ direction  breaks
the longitudinal translational invariance,
and hence allows electrons to make intersideband transitions not to conserve
their longitudinal momenta.~\cite{chu96,tan96}
Moreover, since the transverse electric field is not uniform in
the $y$ direction, the transverse translational invariance is also violated.
Thus the electron-photon scattering processes can include
intersubband transitions. This transmission mechanism is
quite different with the scatterings that induced by a longitudinal
field\,~\cite{tan99} or a time-modulated potential,~\cite{tan96} where
the electrons can only make intrasubband transitions
(the subband index remains unchanged).

The rest of this paper is organized as follows. In Sec. II, the
generalized scattering-matrix method is formulated that has incorporated
a time-dependent mode-matching scheme to solve the time-dependent
Schr\"odinger equation.
The method described in Sec. II is calculated numerically in Sec. III.
From our numerical evidence, we predict that QBS features caused by the
transversely polarized field can occur in narrow constrictions.
Concluding remarks are given in Sec. IV.

\section{Model and Method}

In this section, the coherent inelastic scattering problem in the presence
of a finite-range transverse electric field is formulated. The finite-range
time-dependent electric field is divided by a series of segments,
each of them is described by a $\delta$-profile field.~\cite{chu96}
The matching between these sliced regions has to be performed
in the cascading of the scattering
matrices, from which the transmission and reflection coefficients are
obtained. The conductance $G$ is then expressed in terms of these
coefficients.

Previously, we have investigated transport properties of electrons in
narrow constrictions acted upon by a longitudinally polarized
time-dependent electric field.~\cite{tan99} The potential due to the
longitudinal field that have a finite range in the $x$ direction, but
remain uniform in the $y$ direction. This uniformity in the transverse
direction allows us to propose a matching scheme that avoids slicing the
region covered by the longitudinal time-dependent field. However,
as long as the narrow constriction is acted upon by a transversely
polarized time-dependent field, the translational invariance in the
$y$ direction is breaked --- both intersubband and
intersideband transitions are involved. Hence, we have to develop a
generalized scattering method to formulate the quantum transport problem
when a transverse external field acts upon the narrow constriction.

Since the electric field is
assumed to be applied only on the narrow constriction region, we
need only to formulate this scattering problem in the narrow
constriction region. In the ballistic regime, the length of
the narrow constriction, $L_{c}$, is smaller than
the phase breaking length, $l_{\phi}$,
and hence the electron transport
can be treated as a single particle problem.
Thus the electron transport can be
formulated by a time-dependent Schr\"odinger equation, given by
\begin{equation}
i\hbar\frac{\partial}{\partial t} \, \Psi({\bf x},t) = {\cal H}({\bf
x},t)\,  \Psi({\bf x},t)\, ,
\end{equation}
with the Hamiltonian of the form
\begin{equation}
{\cal H}({\bf x},t) = \left[ {\bf p} + {\frac{e}{c}}{\bf A}({\bf x},t) %
\right]^2 + V_c(y).
\label{eq:ham}
\end{equation}
Here ${\bf p}$ denotes the momentum of an electron
and $V_c(y)$ represents the transverse confinement of the
narrow constriction modeled by a quadratic potential.~\cite{but90}
Taking the Coulomb gauge, the effect of the
transversely polarized electric field can be
represented by a vector potential:
\begin{equation}
{\bf A}({\bf x},t) = -{\frac{c}{\omega}}{\cal E}(x)\sin(\omega t)\hat{y}\, ,
\end{equation}
where ${\cal E}(x)$ represents the profile of the external field with
amplitude ${\cal E}_{0}$ for $|x|<L/2$ and vanishes otherwise.

To be convenient for analysis, below we choose
the length unit $a^{*}=1 / \! k_{{\rm F}}$, the energy unit
$E^{*}=\hbar^2 k_{{\rm F}}^2/(2m^*)$,
the time unit $t^{*}=\hbar / E^{*}$, and field amplitude
${\cal E}_{0}$ in units of $E^{*}/(ea^{*})$, where $-e$ denotes the
electron charge, with effective mass $m^{*}$, and $k_{{\rm F}}$ represents
a typical Fermi wave vector of the reservoir.
Thus we can write the dimensionless transverse confinement
$V_c(y)=\omega_y^2 y^2$, and then gives the
quantized transverse energy levels $\varepsilon_n = (2n+1)\omega_{y}$
and the corresponding wave function $\phi_n(y)$.

The finite-range electric field is divided by $N_{L}$ slices, thus the
width of every slice is given by $\delta L = L/N_{L}$.
Here the sufficiently large
amount of $N_{L}$ is needed such that $\delta L$ is sufficiently
narrow to ensure every slice can be described by a $\delta$-profile
field.~\cite{chu96} The
locations of these $\delta$-profile fields are given by $x_{i}= -L/2 +
(i-1/2)\delta L$, where the positive integer $i=1, 2, \cdots, N_{L}$. By
dividing the profile slice-wisely, the Schr\"odinger equation of the
$i$th $\delta$-profile field is then given by
\begin{widetext}
\begin{eqnarray}
 i {\partial\over\partial t} \Phi^{(i)}({\bf x},t)
= \left[ -\left( {\frac{\partial^2}{\partial x^2}} +
\frac{\partial^2}{\partial y^2} \right) + \omega_{y}^2 y^2 + \left(i
{\frac{ 2{\cal E}_{0}} {\omega}} \frac{\partial}{\partial y}
\sin(\omega t) + \frac{{\cal E}_{0}^2 }{\omega^2} \sin^2(\omega t)
\right) \delta L \delta (x-x_{i})\right] \Phi^{(i)}({\bf x},t)\, .
\end{eqnarray}
Considering a $n$th subband electron incident from left-hand side of
the $i$th $\delta$-profile field, and with incident energy
$\mu^{\prime}$, the
scattering wave function is given by~\cite{chu96}
\begin{mathletters}
\begin{eqnarray}
\Phi_{n}^{(i)}({\bf x},t) &=&
\phi_{n}(y)\exp\left[ ik_n(\mu^{\prime})x - i\mu^{\prime}t \right]
\nonumber \\
&&+ \displaystyle
\sum_{n^{\prime},m^{\prime}}r^{(i)}_{n^{\prime}n}(m^{
\prime})\phi_{n^{\prime}}(y) \exp\left[
-ik_{n^{\prime}}(\mu^{\prime}+ m^{\prime}\omega )x \right]
\exp\left[ -i(\mu^{\prime}+m^{\prime}\omega)t \right] \hspace{5mm}
{\rm if}\ x < x_i , \label{eq:phi1}\\
\Phi_{n}^{(i)}({\bf x},t) &=& \displaystyle
\sum_{n^{\prime},m^{\prime}}t^{(i)}_{n^{\prime}n}(m^{\prime})
\phi_{n^{\prime}}(y) \exp\left[
ik_{n^{\prime}}(\mu^{\prime}+m^{\prime} \omega )x \right] \exp\left[
-i(\mu^{\prime}+m^{\prime}\omega)t \right] \hspace{5mm} {\rm if}\ x
> x_i , \label{eq:phi2}
\end{eqnarray}
\end{mathletters}
\end{widetext}
where the electron is scattered into the subband $n^{\prime}$ and
sideband $m^{\prime}$. The wave vector $k_{n^{\prime}}(\mu^{\prime})
= \sqrt{\mu^{\prime}-\varepsilon_n}$ is the effective wave vector
for the electron with energy $\mu^{\prime}$ and in the $n$th
subband. Here we have defined $\Phi^{(i)}({\bf x},t) =
\sum_n\Phi_n^{(i)}({\bf x},t)$ as a summation over all occupied
incident subbands. The coefficients in Eq.~(\ref{eq:phi1}) and
(\ref{eq:phi2}) have to be determined by the following boundary
conditions:
\begin{equation}
\left.\Phi_n^{(i)}\right|_{x=x_i-\delta} =
\left.\Phi_n^{(i)}\right|_{x=x_i+\delta}\, ,  \label{eq:bound1}
\end{equation}
and
\begin{eqnarray}
&&{\displaystyle \left. {\frac{\partial\Phi_n^{(i)}}{\partial x}}
\right|_{x=x_i+\delta} -\left. {\frac{\partial\Phi_n^{(i)}}{\partial
x}}
\right|_{x=x_i-\delta} } \\
&=& {\displaystyle \left[ i {\frac{ 2{\cal E}_{0}}{%
\omega}} {\frac{\partial}{\partial y}} \sin(\omega t) + {\frac{{\cal
E}_0^2 }{\omega^2 }} \sin^2(\omega t) \right] \delta L
\Phi_n^{(i)}(x=x_i) }\, .\nonumber
\label{eq:bound2}
\end{eqnarray}

\begin{figure}[tbp]
\includegraphics[width=0.41\textwidth,angle=0]{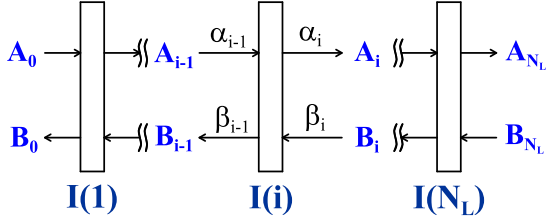}
\caption{ Sketch of the finite-range trnasverse electric field,
which is divided into $N_L$ slices. Every slice is described by a $\protect\delta$%
-profile field, and it is between the $(i-1)$th and the $i$th
region. The coefficients of the right and left going states of the
$i$th region are denoted as ${\bf A}_{i}$ and ${\bf B}_{i}$,
respectively, and these coefficients between successive regions are
connected by the interface matrix ${\bf I}(i)$. $\protect\alpha_{i}$
and $\protect\beta_{i}$ are dummy indices of the $i$th region,
including both subband and sideband indices. } \label{fig:2}
\end{figure}

Imposing the boundary conditions (\ref{eq:bound1}) and (\ref{eq:bound2}) to
perform the matching at all times and given the expression of the matrix
element

\begin{equation}
<l\mid {\frac{\partial}{\partial y}}\mid n^{\prime}> = \sqrt{{\frac{%
\omega_{y}}{2}} }\left[ \sqrt{n^{\prime}}\delta_{l,n^{\prime}-1} - \sqrt{%
n^{\prime}+1} \, \delta_{l,n^{\prime}+1} \right]\, ,
\end{equation}
we obtain the equations relating the reflection coefficients $%
r^{(i)}_{ln}(m) $ and the transmission coefficients $t^{(i)}_{ln}(m)$, 
\begin{mathletters}
\begin{equation}
t^{(i)}_{ln}(m) - r^{(i)}_{ln}(m) = \delta_{m,0} \, \delta_{n,l} \, ,
\label{eq:coeff1}
\end{equation}
and
\begin{eqnarray}
 &&\delta_{m,0} \ \delta_{n,l} \, k_{n}(\mu')   \nonumber \\
&=& k_{l}(\mu^{\prime}+m\omega) \left[ r^{(i)}_{ln}(m) + t^{(i)}_{ln}(m) %
\right]  \nonumber \\
&&+ i\frac{{\cal E}_0}{\omega}\delta L\, \sum_{n^{\prime},
m^{\prime}} \left[\delta_{m^{\prime}, m+1} -
\delta_{m^{\prime}, m-1}\right] \nonumber \\
&&\hspace{20mm}  \times <l\mid \frac{\partial}{\partial y}\mid
n^{\prime}>
t^{(i)}_{n^{\prime}n}(m^{\prime})  \\
&&+ i\frac{{\cal E}_0^2}{4\omega^2}\delta L\, \left[
2t^{(i)}_{ln}(m) + t^{(i)}_{ln}(m+2) + t^{(i)}_{ln}(m-2) \right]
.\nonumber
 \label{eq:coeff2}
\end{eqnarray}
From these expressions associated with the wave vector, $k_n(\mu)$, along
the channel direction, it is clear that the time-dependent field-induced
electron transitions do not conserve the longitudinal momenta. This means
when the time-dependent field has a finite longitudinal profile
(excluding the reservoirs), the possibility of these transition processes can be made. In
Eq.~(\ref{eq:coeff2}), we can see that the ${\cal E}_0$ term causes the
intersubband and intersideband transitions by emitting or absorbing
one $\hbar\omega$ (one-photon process), while the
${\cal E}_0^2$ term contributes
to the intersubband and intersideband transitions by
emitting or absorbing two $\hbar\omega$.
Solving Eqs.~(\ref{eq:coeff1}) and (\ref{eq:coeff2}), we obtain the
transmission coefficients $t^{(i)}_{ln}(m)$ and reflection coefficients $%
r^{(i)}_{ln}(m)$ of the $i$th slice. For an electron incident from the
right-hand side of the $i$th slice in the same subband, $n$, and energy, $%
\mu^{\prime}$, the transmission coefficient $\widetilde{t}^{(i)}_{ln}(m)$
and the reflection coefficient $\widetilde{r}^{(i)}_{ln}(m)$ differ from
those for an electron incident from the left-hand side of the $i$th slice
only by a phase factor of unit modulus, given by
\begin{mathletters}
\begin{equation}
\widetilde{t}^{(i)}_{ln}(m) = t^{(i)}_{ln}(m)\, \exp\left\{ 2 i \left[
k_{l}(\mu^{\prime}+m\omega) -k_{n}(\mu^{\prime})\right] x_i\right\}\, ,
\end{equation}
and
\begin{equation}
\widetilde{r}^{(i)}_{ln}(m) = r^{(i)}_{ln}(m)\, \exp\left\{-2 i \left[
k_{l}(\mu^{\prime}+m\omega) + k_{n}(\mu^{\prime})\right] x_i\right\} \, .
\end{equation}
\end{mathletters}

In general, for an electron incident from the left-hand side of the $i$th
slice (the $(i-1)$th region) in the subband $n_{i-1}$ and at energy $\mu +
m_{i-1}\omega$, this incident state is denoted as $\alpha_{i-1} =
(n_{i-1},m_{i-1})$. The electron may be transmitted to the right-hand side
of the $i$th slice ($i$th region) into the state $\alpha_i= (n_i,m_i)$ with
a transmission coefficient $t_{\alpha_{i},\alpha_{i-1}}$. Also, the electron
may be reflected to the left-hand side of the $i$th slice ($(i-1)$th region)
into the state $\beta_{i-1}= (n^{\prime}_{i-1},m^{\prime}_{i-1})$ with a
reflection coefficient $r_{\beta_{i-1},\alpha_{i-1}}$. Similarly, for an
electron incident from th right-hand side of the $i$th slice in the incident
state $\beta_i=(n^{\prime}_i,m^{\prime}_i)$, the corresponding transmission
coefficient and reflection coefficient due to this slice are given by $%
\widetilde{t}_{\beta_{i-1},\beta_{i}}$ and $\widetilde{r}_{\beta_{i-1},%
\beta_{i}}$, respectively. After defining these coefficients, we can
establish the scattering matrix equation, given by
\end{mathletters}
\begin{equation}
\left[
\begin{array}{c}
{\bf A}_{i} \\
{\bf B}_{i-1}
\end{array}
\right] = {\bf S}(i-1,i) \left[
\begin{array}{c}
{\bf A}_{i-1} \\
{\bf B}_{i}
\end{array}
\right]\, ,
\label{eq:smat1}
\end{equation}
where ${\bf A}_{i}$ and ${\bf B}_{i}$ are the coefficients of the right- and
the left-going states in the $i$th region, respectively, as illustrated in
Fig.~\ref{fig:2}. Here ${\bf S}(i-1,i)$ is the scattering matrix which
connects the $(i-1)$th to the $i$th region, across the $i$th slice, given by
\begin{equation}
{\bf S}(i-1,i) = \left[
\begin{array}{cc}
{\bf t}(i) & {\bf \widetilde{r}}(i) \\
{\bf r}(i) & {\bf \widetilde{t}}(i)
\end{array}
\right]\, .
\end{equation}
Here, ${\bf t}(i)$ and ${\bf r}(i)$ denote the transmission and
reflection matrices of the right-going electron at the $i$th slice,
respectively, and the tilded two refer to the contribution of
left-going electron. Matching between these sliced regions has to
be performed in the cascading of the scattering matrices. We should point
out the reason for using the scattering matrix formalism, instead of the
transfer matrix method, is to avoid the use of truncation schemes required
in dealing with the exponentially growing solutions. The scattering matrix
method is stable and accurate without any special treatment for any of the
intermediate states.

To do the piece-wise matching, we start from rearranging Eq.~(\ref{eq:smat1}%
) to obtain the following matrix equation
\begin{equation}
\left[
\begin{array}{c}
{\bf A}_{i-1} \\
{\bf B}_{i-1}
\end{array}
\right] = {\bf I}(i) \left[
\begin{array}{c}
{\bf A}_{i} \\
{\bf B}_{i}
\end{array}
\right]\, ,  \label{eq:i-mat}
\end{equation}
which connects the coefficients of the successive regions across the $i$th
slice. Here ${\bf I}(i)$ is the interface matrix of the $i$th slice,
defined by
\begin{equation}
{\bf I}(i) = \left[
\begin{array}{cc}
{\bf I}_{11}(i) & {\bf I}_{12}(i) \vspace{1mm} \\
{\bf I}_{21}(i) & {\bf I}_{22}(i)
\end{array}
\right]\, ,
\end{equation}
in which
\begin{eqnarray}
{\bf I}_{11}(i) &=& {\bf t}(i)^{-1} ,
\nonumber \\
{\bf I}_{12}(i) &=& -{\bf t}(i)^{-1}
{\bf \widetilde{r}}(i) ,  \nonumber \\
{\bf I}_{21}(i) &=& {\bf r}(i) {\bf t}(i)^{-1} ,  \nonumber \\
{\bf I}_{22}(i) &=& {\bf \widetilde{t}}(i)
-{\bf r}(i){\bf t}(i)^{-1} {\bf \widetilde{r}}(i) .
\end{eqnarray}
In general, for the regions up to the ($i-1$)th slice, we have
\begin{equation}
\left[
\begin{array}{c}
{\bf A}_{i-1} \\
{\bf B}_{0}
\end{array}
\right] = {\bf S}(0,i-1) \left[
\begin{array}{c}
{\bf A}_{0} \\
{\bf B}_{i-1}
\end{array}
\right]\, ,  \label{eq:s-mat2}
\end{equation}
where ${\bf S}(0,i-1)$ is the scattering matrix connecting the $0$th region
to the $(i-1)$th region, defined by
\begin{equation}
{\bf S}(0,i-1) = \left[
\begin{array}{cc}
{\bf S}_{11}(0,i-1) & {\bf S}_{12}(0,i-1) \vspace{1mm} \\
{\bf S}_{21}(0,i-1) & {\bf S}_{22}(0,i-1)
\end{array}
\right]\, .
\end{equation}
Imposing Eqs.~(\ref{eq:i-mat}) and (\ref{eq:s-mat2}), the coefficients ${\bf %
A}_{i-1}$ and ${\bf B}_{i-1}$ may be eliminated, and then we obtain the
matrix equation connecting the $0$th to the $i$th region, given by
\begin{equation}
\left[
\begin{array}{c}
{\bf A}_{i} \\
{\bf B}_{0}
\end{array}
\right] = {\bf S}(0,i) \left[
\begin{array}{c}
{\bf A}_{0} \\
{\bf B}_{i}
\end{array}
\right]\, .
\end{equation}
The submatrices of this scattering matrix ${\bf S}(0,i)$ are,
explicitly,\cite{ko88}
\begin{widetext}
\begin{eqnarray}
{\bf S}_{11}(0,i) &=& \left[ {\bf I}_{11}(i) - {\bf S}_{12}(0,i-1) {\bf I}%
_{21}(i) \right]^{-1} {\bf S}_{11}(0,i-1)\, ,  \nonumber \\
{\bf S}_{12}(0,i) &=& \left[ {\bf I}_{11}(i) - {\bf S}_{12}(0,i-1) {\bf I}%
_{21}(i) \right]^{-1} \, \left[ {\bf S}_{12}(0,i-1) {\bf I}_{22}(i)
- {\bf I}_{12}(i) \right]\, ,  \nonumber \\
{\bf S}_{21}(0,i) &=& {\bf S}_{21}(0,i-1) + {\bf S}_{22}(0,i-1) {\bf I}%
_{21}(i) {\bf S}_{11}(0,i)\, ,  \nonumber \\
{\bf S}_{22}(0,i) &=& {\bf S}_{22}(0,i-1) {\bf I}_{22}(i) + {\bf S}%
_{22}(0,i-1) {\bf I}_{21}(i) {\bf S}_{12}(0,i)\, .
\label{eq:s(0,i)}
\end{eqnarray}
\end{widetext}
This iterative procedure is not as easy to evaluate in terms of the
transfer-matrix method, which simply inverses a product of matrices. More
precisely, once the system is acted upon by an external time-modulated
field, the evanescent modes play an important role due to inelastic
scatterings. In this situation, we may prefer to use the scattering-matrix
method to gain the stability for the numerical computation.

By iterating Eq.~(\ref{eq:s(0,i)}), we obtain the scattering matrix ${\bf S}%
(0,N_{L})$ which satisfies the matrix equation:
\begin{equation}
\left[
\begin{array}{c}
{\bf A}_{N_{L}} \\
{\bf B}_{0}
\end{array}
\right] = {\bf S}(0,N_{L}) \left[
\begin{array}{c}
{\bf A}_{0} \\
{\bf B}_{N_{L}}
\end{array}
\right]\, .
\end{equation}
This equation describes the electron transport through the whole
time-modulated region. The incident state is assumed to be $\alpha_{{\rm in}%
}=(n_{0},0)$ such that the elements of the incident coefficient ${\bf A}_{0}$
can be expressed as $\delta_{n,n_{0}}\delta_{m,0}$, i.e., only $(n_{0},0)$
is the nonzero element. Setting ${\bf B}_{N_{L}}={\bf 0}$, we have
\begin{mathletters}
\begin{equation}
{\bf A}_{N_{L}} = {\bf S}_{11}(0,N_{L}) {\bf A}_{0}\, ,
\end{equation}
and
\begin{equation}
{\bf B}_{0} = {\bf S}_{21}(0,N_{L}) {\bf A}_{0}\, .
\end{equation}
The transmission coefficient for an electron incident from the initial state
$\alpha_{{\rm in}}=(n_{0},0)$ and transmitted, by the finite-range
transverse field, into the final state $\alpha_{{\rm f}}=(n_{{\rm f}},m_{%
{\rm f}})$ is denoted by $t_{\alpha_{{\rm f}},\alpha_{{\rm in}}} =
(A_{N_{L}})_{\alpha_{{\rm f}}}$, where $(A_{N_{L}})_{\alpha_{{\rm f}}}$ is
an element of ${\bf A}_{N_{L}}$. The current transmission coefficient,
corresponding to this inelastic scattering process, is then given by
\end{mathletters}
\begin{equation}
T_{\alpha_{{\rm in}}}^{\alpha_{{\rm f}}}=\left[ {\frac{ k_{n_{{\rm f}}}(\mu
+m_{{\rm f}}\omega)}{k_{n_{0}}(\mu) }}\right] \left| t_{\alpha_{{\rm f}%
},\alpha_{{\rm in}}} \right|^2\, .
\end{equation}
Therefore, the zero temperature conductance may obtained by summing over all
the possible incident and transmitted states, given by
\begin{equation}
G = {\frac{2e^2}{h}}\sum_{\alpha_{{\rm in}}} \sum_{\alpha_{{\rm f}}}\,
T_{\alpha_{{\rm in}}}^{\alpha_{{\rm f}}} = {\frac{2e^2}{h}} \sum_{\alpha_{%
{\rm in}}}\, T_{\alpha_{{\rm in}}}\, ,
\end{equation}
where $T_{\alpha_{{\rm in}}}$ is the current transmission coefficient from
the incident state $\alpha_{{\rm in}}$. Since the incident sideband is
specified to be $m=0$ for an arbitrary incident subband $n_{0}$, so that the
summation $\sum_{\alpha_{{\rm in}}} = \sum_{n_{0}=0}^{N}$, where $N+1$ is
the number of propagating subbands for the chemical potential $\mu$. But for
the final states, both the subband and sideband indices are arbitrary, thus $%
\sum_{\alpha_{{\rm f}}} = \sum_{n_{{\rm f}}=0}^{N} {\sum^{\prime}}_{m_{{\rm f%
}}} $ is expected to be a double sum. Here the superscript prime indicates
that summation is over $m_{{\rm f}}$ such that $k_{n_{{\rm f}}}(\mu +m_{{\rm %
f}}\omega)$ is real, namely that only occupied subbands are included for the
scattering states. The conservation of current, given by the condition
\begin{equation}
\sum_{\alpha_{{\rm f}}}\, \frac{k_{n_{{\rm f}}}(\mu+m_{{\rm f}}\omega)}{%
k_{n_{0}}(\mu)} \, \left[ \, \left| t_{\alpha_{{\rm f}},\alpha_{{\rm in}}}
\right|^2 + \left| r_{\alpha_{{\rm f}},\alpha_{{\rm in}}} \right|^2 \,
\right] = 1\, .
\end{equation}
is used to check our numerical accuracy.

\section{Numerical Results and Discussion}

\begin{figure}[tbp]
\includegraphics[width=0.41\textwidth,angle=0]{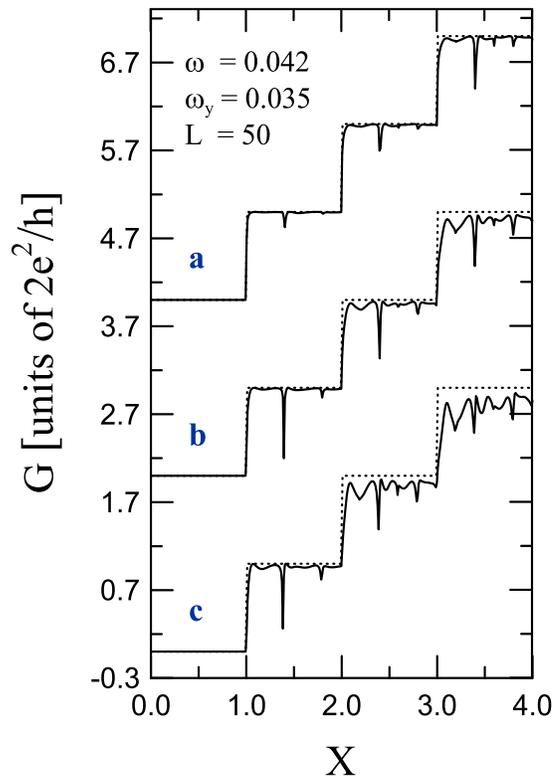}
\caption{ Conductance $G$ as a function of $X$ for frequency $\protect\omega %
= 0.042$ ($\simeq 0.6\Delta\protect\varepsilon$), with time-modulated range $%
L = 50$. The amplitude of the electric field are ${\cal E}_{0}$ = a, 0.002 ($%
\simeq 22.6$ V/cm); b, 0.003 ($\simeq 33.9$ V/cm); and c, 0.004
($\simeq 45.2 $ V/cm). The curves are vertically offset for clarity.
} \label{fig:4}
\end{figure}
\begin{figure}[tbp]
\includegraphics[width=0.41\textwidth,angle=0]{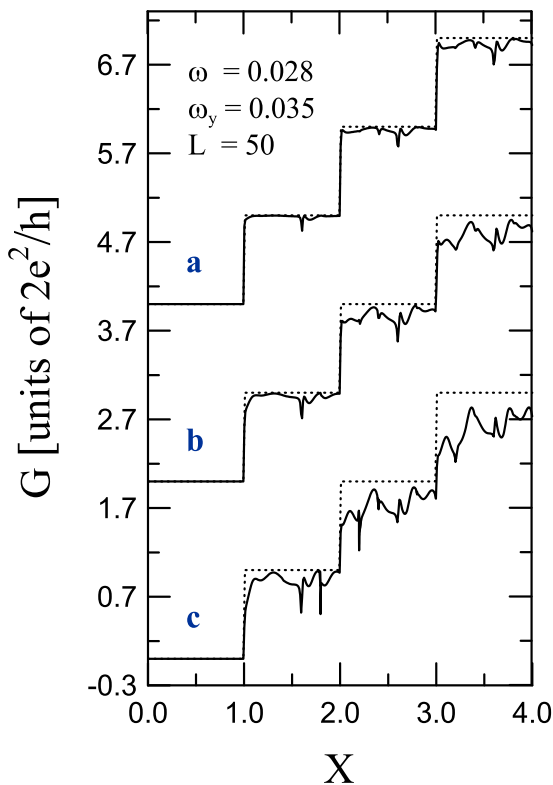}
\caption{ Conductance $G$ as a function of $X$ for frequency $\protect\omega %
= 0.028$ ($\simeq 0.4\Delta\protect\varepsilon$), with time-modulated range $%
L = 50$. The amplitude of the electric field are ${\cal E}_{0}$ = a, 0.002 ($%
\simeq 22.6$ V/cm); b, 0.003 ($\simeq 33.9$ V/cm); and c, 0.004
($\simeq 45.2 $ V/cm). The curves are vertically offset for clarity.
} \label{fig:3}
\end{figure}

In this section the behavior of the conductance $G$ is studied. To
facilitate the experimental performance we fix the length $L$ of the
time-modulated region while varying the field strength ${\cal
E}_{0}$, in
which the angular frequencies are chosen to be $\omega=0.028$ ($%
\nu=\omega/2\pi\cong 61\, {\rm GHz}$) and 0.042 ($\nu\cong 91\, {\rm GHz}$),
as depicted in Figs.~\ref{fig:3} and \ref{fig:4}, respectively. The $L$ in
both the figures are chosen to be $L=50\ (\simeq 0.4\ \mu{\rm m})$. The $G$
characteristics are represented by the dependence on $X$, the suitably
rescaled chemical potential $\mu$. According to this scale, when $\mu$ is
changed by a subband energy level spacing $\Delta\varepsilon$, it
corresponds to $\Delta X=1$, and when $\mu$ is changed by $\hbar\omega$, it
corresponds to $\Delta X=\omega / \Delta\varepsilon = 0.4\ {\rm and}\ 0.6$
for Figs.~\ref{fig:3} and \ref{fig:4}, respectively. In addition, when $X=N$%
, $\mu$ is at the $N$-th subband bottom, namely $N=1,\ 2,\ {\rm or}\ 3$ in
these figures.

In our numerical examples, the narrow constriction is chosen to be that in a
high mobility
${{\rm GaAs-Al}_{x}{\rm Ga}_{1-x}{\rm As}}$ with a typical electron density
$n \sim 2.5 \times 10^{11}$ cm$^{-2}$, and $m^{*} = 0.067 m_{e}$.
Correspondingly, our choice of length unit $a^{*} = 1/k_{{\rm F}} = 79.6$
\AA, energy unit $E^{*} = \hbar^{2}k_{{\rm F}}^{2}/(2m^{*}) = 9$ meV, and
frequency unit $\omega^{*} = E^{*}/\hbar = 13.6$ THz. We also choose
$\omega_y = 0.035$ such that the transverse energy level spacing
$\Delta\varepsilon = 2\omega_{y}=0.07$,
and the effective narrow constriction
width is of the order of $0.1\ \mu{\rm m}$. In the following, in presenting
the dependence of $G$ on $\mu$, it is more convenient to plot $G$ as a
function of $X$ instead, where
\begin{equation}
X = {\frac{\mu}{\Delta\varepsilon}} + {\frac{1}{2}}\, .
\end{equation}
With this conversion, $X$ is in units of $\Delta\varepsilon$, and the
integral value of $X$ is the number of propagating channels through the NC.

In Fig.~\ref{fig:3}, the field amplitudes are chosen to be
${\cal E}_0=0.002,\ 0.003,\ {\rm and}\ 0.004$ for Figs.~\ref{fig:3}a-c,
respectively. The angular frequency $\omega$ is chosen to be 0.028, whose
energy interval $\omega$ corresponds to an interval $\Delta
X=\omega/\Delta\varepsilon =0.4$. The dotted curves are the unperturbed
results. In general, we find the suppressed features in $G$ that escalate
with both the chemical potential and ${\cal E}_0$, as depicted in Figs.~%
\ref{fig:3}a-c. Besides, there are dip structures in these figures, which can
be understood to be the formation of QBS's due to time-modulated
fields.~\cite{chu96,bag92,tan96,tan97}
These QBS's, formed at energies near the threshold of
subbands, trap temporarily conduction
electrons and give rise to drops in $G$.
However, the trapped electron can also be excited out of the QBS,
resulting a smaller $G$ reduction: $|G|< 1$, in units of
$2e^2/h$.~\cite{chu96} In contrast,
the impurity-induced dips are the results of merely
elastic scattering and thus have $G$ reduction $|G|=1$. \cite{chu89,bag90}
It is shown when we increase the ${\cal E}_{0}$,
these dip structures become
more deeper and shift slightly toward the lower energy direction. Our
results demonstrate the manifestation of a new, and time-dependent
electric-field-induced QBS formed from
transitions due to coherent inelastic scatterings.

Figures 3a-c have common types of dip structures. First, the dip structures
are found at around $X$ = 1.6, 2.6, and 3.6, that is, at $\Delta X=0.4$
beneath a subband bottom. These dip structures are due to the processes that
an electron in the $N$th subband, and at energy $(N+1)-\Delta X$, can absorb
an energy $\hbar\omega$ and become bounded in the QBS just beneath the
threshold of the $(N+1)$th subband. In other words, this process is $(\Delta
n,\Delta m) = (+1,+1)$. Second, the small dip structures at around $X$ = 2.4
and 3.4 are attributed to the processes that an electron in the $N$th subband, and
at energy $N+\Delta X$, can give away $\hbar\omega$ to be trapped
temporarily in the $(N+1)$th subband bottom. This process corresponds to
$(\Delta n,\Delta m) = (+1,-1)$. Third, the dips at around $X$ = 2.2 and 3.2
are contributed by two different kinds of transitions, that is, (+2,+2) and
(-1,+3) processes. The former corresponds to that an electron in the $N$th
subband with energy $(N+2)-2\Delta X$ can absorb $2\hbar\omega$ to the
$(N+2)$th subband bottom, and the latter corresponds to those make
transitions from the $N$th subband, with energy $(N-1)+3\Delta X$, to the
$(N-1)$th subband bottom. We should point out that all the above are
intersubband and intersideband transitions.

Moreover, in Figs.~\ref{fig:3}b and c, there are dip structures at around $X$
= 3.0 and 4.0, which correspond to (+2,0) transitions. These structures are
due to intersubband and intrasideband transitions, namely that the electron
energy remains unchanged. Besides, in Fig.~\ref{fig:3}c, there are dips at
around $X$ = 1.8, 2.8, and 3.8, which correspond to (0,-2) transitions.
We can see that these multi-photon transitions manifest only
when the applied field intensity is larger.

In Fig.~\ref{fig:4}, the field amplitudes are chosen to be the same as
Fig.~\ref{fig:3}. The angular frequency $\omega$
is chosen to be 0.042, whose
energy interval $\omega$ corresponds to an interval
$\Delta X=\omega/\Delta\varepsilon =0.6$.
There are dip structures in these figures
associated with the formation of time-depdent field-induced
QBS's.~\cite{chu96,bag92,tan96,tan97}
There are common types of dip structures in Figs.~\ref{fig:4}a-c.
First, the dip structures are found at around $X$ = 1.4,
2.4, and 3.4, which are associated with (+1,+1) processes. These dip
structures correspond to that an electron in the $N$th subband, and at
energy $(N+1)-\Delta X$, can absorb an energy $\hbar\omega$ and become
bounded as a QBS in the $(N+1)$th subband. Second, the small dips at around
$X$ = 2.6 and 3.6, associated with (+1,-1) processes, correspond to that
an electron in the $N$th subband with energy $N+\Delta X$ can give away
$\hbar\omega$ to be trapped temporarily in the $(N+1)$th subband bottom.
Third, the dips at around $X$ = 1.8, 2.8, and 3.8, associated with (+2,+2) processes,
correspond to that an electron in the $N$th subband, at energy
$(N+2)-2\Delta$, can absorb $2\hbar\omega$ to the $(N+2)$th subband bottom.
Moreover, in Figs.~\ref{fig:4}b and c, dip structures at around
$X$ = 3.0 and 4.0 are identified to be (+2,0) intersubband and
intrasideband transitions. Furthermore, in Fig.~\ref{fig:4}c, there are
dips at around $X$ = 2.2 and 3.2, which are associated with (0,-2) transitions.

As mentioned in our previous work,~\cite{tan99} to observe the above
predicted effects, the experimental setup needs to fulfill two requirements.
First, the bolometric
effect due to the absorption of photons in the QPC's end-electrodes has to
be suppressed or totally eliminated. Recent experiments show that the
transport characteristics are masked by the bolometric effect when the
entire QPC, including the end-electrodes, is exposed to the incident
electromagnetic field.~\cite{ala98} Second,
the length $L$ of the time-dependent field has to be
shorter than the wave length of the incident field. The purpose is to
increase the coupling between the electrons and the photons by breaking the
longitudinal translational invariance. That the coupling between the photon
field and the conduction electrons can be much enhanced, when either the
electrons are confined or the time-dependent field has a localized profile,
has been pointed out recently by Yakubo {\it et al.\/}~\cite{yak96}
Thus the QPC needs to be in the near-field regime of the
time-dependent field.

To avoid the bolometric effect, we suggest
to apply ac field to the split gates of the QPC instead of shining
an electromagnetic wave upon the QPC. The split gates are negatively biased
with respect to a common ground, and made of
superconducting materials with superconducting wires connecting to
an ac-signal generator.
This generator can be available using the IMPATT
diode that has successfully been demonstrated to cover the complete millimeter
range (30-300 GHz).~\cite{bha84}
This proposed experimental setup is expected to generate
a transversely polarized electric field only in the narrow
constriction region while keeping the two-end electrodes
from the time-modulated field.  In this work,
though the time-dependent region covers only part of the narrow
constriction, we believe these two situations will manifest similar
features.
Given the availability of millimeter wave sources,~\cite{bha84} the
suggested experimental setup would be manageable by the present
nanotechnology. The features reported in this work, however, are
not limited to millimeter waves.

\section{Conclusion}

A generalized scattering-matrix method has been developed for investigating
coherent quantum transport in narrow constrictions with a transversely
polarized time-dependent electric field. This method allows us to
solve nonpertubatively the time-dependent Schr\"odinger equation in the
numerical sense. Since the energy conservation law is violated in
such a time-modulated system, a conventional transfer-matrix method
technique is inapplicable. Using the present numerical method, not only
the transmission and reflection probabilities of systems can be
calculated, all the subband and sideband states can be
obtained.

The scattering processes due to the time-dependent external field are
both inelastic and coherent. Since this field is transversely
polarized, electrons can make both intersubband and intersideband
transitions. This increases the complexity in calculation, but has
more interesting features.
Different dip structures
associated with different intersubband and intersideband transitions to
the vicinity of a subband bottom are found.
These dip structures can be understood as
the formation of a QBS at energy near a subband bottom due to its singular
DOS.~\cite{chu96}
Moreover, due to the tunability of frequency and
intensity of the field, this proposed configuration can be
applied to be a high-frequency detector.
We expect that these dip structures could also be found
when the QPC has a varying width. We hope that the present method will
be utilized to study new transport phenomena in mesoscopic nanostructures.

\begin{acknowledgments}
The authors would like to thank the National Science Council of the
Republic of China for financially supporting this research under
Contract No. NSC88-2112-M-009-028. Computational facilities
supported by the National Center for High-performance Computing are
gratefully acknowledged.
\end{acknowledgments}

\end{document}